\documentstyle[11pt,newpasp,twoside,epsf]{article}
\reversemarginpar

\begin{document}
\title{Inhomogeneous Metallicities and Cluster Cooling Flows}
\author{R. Glenn Morris and Andrew C. Fabian}
\affil{Institute of Astronomy, Madingley Road, Cambridge CB3 0HA, UK.}

\begin{abstract}
  The theme of this contribution is the idea that the metals in the
  intracluster medium (ICM) may be inhomogeneously distributed on small
  scales. We focus on the possible interplay between this and the cooling
  flow process. We begin with a brief review of some evidence for a patchy
  distribution of metals in the ICM\@. Next we outline a simple numerical
  model for this within the framework of the cooling flow hypothesis.
  Finally we present some of our recent results on the presence of
  gradients in the observed abundance profiles.
\end{abstract}

\section{Introduction}

\subsection{The Intracluster Metallicity}

It has been known for some time that the mean metallicity of the
intracluster medium (ICM) is about 0.3 solar (hereinafter $Z_{\odot}$);
both for nearby (Edge \& Stewart 1991), and distant (Mushotzky \&
Loewenstein 1997) clusters. It is probably fair to say that there is no
{\it a priori} reason why the ICM should be uniformly enriched to this
value; for example, if we just consider the relative mass in gas and stars
today, which can be $\sim$ 5:1.

In addition, simple calculations show that the diffusion time of ions in
the ICM is extremely long. For example, Ezawa et~al.\ (1997) estimate that
iron can only diffuse $\sim 10$\,kpc or so in a Hubble time. Of course,
other effects such as convection, mergers, etc.\ will act to smooth out the
metal profile. These are complicated issues, but it does not seem
unreasonable to adopt as a working hypothesis the idea that the metal
distribution might not be homogeneous and see what the consequences might
be.

Several recent X-ray images of clusters appear to show ``cold-fronts''
where the surface brightness changes abruptly; for example, the Markevitch
et~al.\ (2000) image of A2142. This was used by Ettori \& Fabian (2000) to show
that thermal conduction must be reduced by a factor of 250--2500 compared
to the ``classical'' value. Here we will only add the comment that if
thermal conduction is suppressed, then ion movement will be even more so,
since for significant thermal conduction to take place individual electrons
do not actually have to travel that far.

Analysis of the {\it Chandra} image of 4C+55.16 (a powerful radio source in
a cooling flow cluster at $z = 0.24$) by Iwasawa et~al.\ (in preparation)
shows an extremely high abundance (super-solar) within the central 50\,kpc,
dropping to around $0.5\,Z_{\odot}$ outside this radius.

It is important to stress that these are {\it not} the kind of
inhomogeneity we will consider in the rest of this contribution. We wish to
explore the idea that the ICM metal distribution might vary on {\it small},
essentially unresolved scales. The previous two points are merely evidence
that: (i) regions of very different properties can co-exist in the ICM at
essentially sharp boundaries; and (ii) regions of extremely high
metallicity may exist in the ICM.

\subsection{Cooling Flows}

Recent observations with the {\it XMM-Newton} {\sc rgs} show a discrepancy
with the predictions of the standard cooling flow model, with an absence of
some of the expected lines from low-temperature species. Various possible
explanations have been suggested for this apparent conflict: heating;
mixing; differential absorption; metallicity variations; etc. It was for
this reason that we first began to consider the last of these, namely
small-scale metallicity variations, but from this starting point our
investigations have subsequently developed along different lines.

\section{The Model}

\subsection{System Specification and Evolution}

We assume spherical symmetry so that the system can be described by a
one-dimensional model. A cluster is modelled as a two-component system
comprised of hot gas in the potential well of a dark matter halo. The
latter is represented by a standard NFW profile. In order to create a
realistic cluster mass profile, we make use of the calibrated virial
scaling relations of Evrard, Metzler, \& Navarro (1996), and take a
concentration parameter of 5 as typical of the cluster regime.

To date, the gas has been modelled only as a single-phase medium, i.e.\
only one temperature and density at any given radius. For initial
conditions we assume a hydrostatic pressure profile of the isothermal form.
The outer boundary pressure is adjusted so that the gas mass is some
appropriate fraction $(\approx 0.2)$ of the total mass.

The gas is allowed to evolve under the effects of gravity and radiative
energy loss, using a constant-pressure outer-boundary condition. The 1D
hydrodynamics equations are discretised using the Lagrangian method of
Thomas (1988). Put simply, the radial range is divided into a large number
of zones which are then evolved forwards in discrete time-steps, using an
adaptive step-size to maintain stability.

\subsection{Metallicity}

The radiation from the gas is modelled using the {\sc mekal} spectral code.
Consequently, the metallicity is a relevant parameter. Each radial zone may
have a different set of abundances for the 15 elements included in {\sc
mekal}. Initially, we integrate spectra to build up a discrete evaluation
of the cooling function. This is strongly dependent upon the metallicity,
with different species controlling the cooling in different temperature
regimes. At high ($\ga 10^{8}$\,K) temperatures, the cooling is dominated
by the H, He continuum; whereas at lower temperatures line cooling from
heavy elements dominates (mainly iron in the $10^{6}$--$10^{7}$\,K regime,
with oxygen significant at lower temperatures).

The cooling function may itself be integrated in order to obtain a value
for the cooling time. In practice, we only follow the gas cooling down as
far as $10^{5}$\,K. At such temperatures, the cooling time is becoming
prohibitively short for continued computation. Furthermore, the gas has
left the X-ray regime and may therefore be neglected (for our purposes). In
practice, if any zone falls below $10^{5}$\,K we remove it from subsequent
calculations and allow the remainder of the gas to adiabatically expand to
fill the space (a crude representation of the rapid collapse of cold gas
into condensed objects of negligible volume).

\subsection{Spectra}

At given times during the evolution of the system, X-ray spectra are
produced. In order to facilitate comparison with observations, we first
integrate along a cylindrical line of sight through the cluster at a given
projected radius, to calculate a quantity that may be referred to as the
``spectral surface brightness''. Integrating again over a range of
projected radii leads to the spectrum due to an annulus. The spectra are
then redshifted in terms of energy and scaled according to the luminosity
distance. Data are converted to {\it XSPEC} table-model format {\sc fits}
files. Simulated observations may then be produced, by convolving the model
spectra with the response function of a particular detector (we make use of
the {\it Chandra} {\sc acis-s} proposal-planning response files). The
effects of an absorbing column $N_{\rm H}$ are added by multiplying the
model spectra with an {\it XSPEC} {\tt phabs} model.

\section{Results}

\begin{figure}
  \plotfiddle{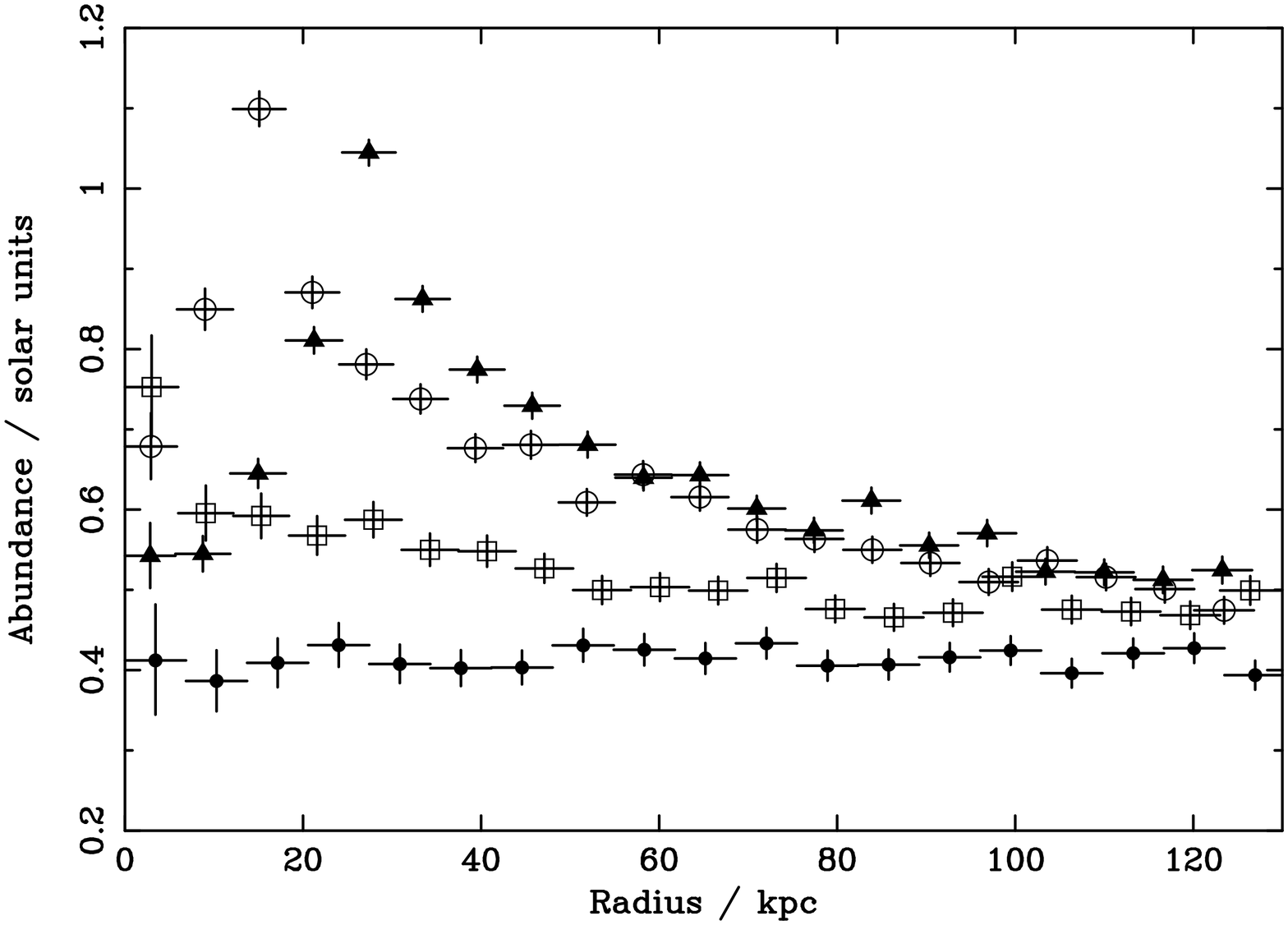}{235pt}{0}{50}{50}{-200}{280}
  \caption{Time evolution of the abundance profile for a bimodal
    distribution of metallicities. Filled circles - 0.0\,Gyr; open squares
    - 0.6; open circles - 0.9; filled triangles 1.0. Vertical error bars
    are 1$\sigma$.}
\end{figure}

To investigate the effects of an inhomogeneous metal distribution, we have
begun by considering one of the simplest possible inhomogeneities: 9 out of
every 10 radial zones are made to be a pure H, He plasma; whilst the tenth
zone has a 5\,$Z_{\odot}$ abundance of metals. This is clearly a very
simple parameterization, but we are concerned not with making detailed
predictions, but rather with seeing what the general trends of behaviour
might be. In this section we illustrate some of the results for a cluster
with the following properties: virial temperature 8.6\,keV; gas fraction
0.17; $z = 0.017$; $N_{\rm H} = 10^{21}\,{\rm cm}^{-2}$; simulated
observation time 25\,ks.

Figure~1 shows the results of fitting the fake observations (in the range
0.3--7\,keV) for each annulus at various times in the evolution of the
system, using a single temperature {\tt mekal} model (multiplied by a {\tt
phabs} component). We fix $N_{\rm H}$ and $z$ at the correct values and
allow the temperature, abundance and normalization of the {\tt mekal}
component to vary freely.

\subsection{Abundance Gradients}

Note firstly that the initial conditions are fit as a uniform abundance
profile, despite the strong variations in metallicity that are actually
present on small scales. This is since in the coronal equilibrium
approximation the ions radiate in exactly the same way whether they are
concentrated in a particular region or uniformly dispersed throughout the
emitting volume. Thus observationally any fluctuations in abundance that
occur on small spatial scales may be very difficult to detect.

As the system evolves, more interesting effects are produced. Initially,
there is a rise in abundance in the central regions, leading to the
appearance of an abundance gradient in the cluster. As further time passes,
the abundance in the very centre of the cluster begins to drop, so that an
abundance peak is produced at an off-centre position. The peak works
outwards with time. It is important to stress that in these simulations,
the actual abundance of the modelled ICM is not time-dependent, but remains
fixed. A control simulation with a homogeneous distribution of metals shows
no such effects - the abundance profile remains flat.

The apparent rise in abundance in the central regions may be explained as
follows. The bulk of the continuum radiation in these models is supplied by
the metal-poor gas; whereas all the line emission is due to the metal-rich
gas. This has a significantly shorter cooling time than the metal-poor gas
owing to its enhanced radiation at all temperatures. The metal-poor gas
therefore cools relatively slowly, and consequently we can regard the
strength of the continuum as being fairly constant. The high abundance
zones on the other hand cool much more rapidly, and as they do so they
enter the temperature regime where line emission is even more significant
in the cooling function. The net result is that the strength of the
emission lines relative to the continuum increases with time, due to the
differential cooling rates of the gas contributing to these two components.
Recall that it is precisely from the strength of emission lines compared to
the continuum (i.e.\ the equivalent width) that abundances are determined.
Thus the behaviour outlined above is fit as an increase in abundance. The
effects are obviously greatest in the central regions where the density is
highest and the two-body brem{\ss}trahlung process is most intense. The
drop in the central abundance at late times occurs as metal-rich gas cools
out and is lost from the X-ray wave-band.

This is an interesting result, because abundance gradients of this form
have been detected in several clusters, for example with {\it ASCA} in
Centaurus (Allen \& Fabian 1994) and AWM 7 (Ezawa et~al.\ 1997). The
presence of a gradient appears to be correlated with the presence of a
cooling flow (e.g., De~Grandi \& Molendi 2001).

Recent {\it Chandra} observations of Centaurus (Sanders \& Fabian, in
preparation) and A2199 (Johnstone et~al., in preparation) seem to show
abundance profiles very similar to those produced here, with a gradient in
the outer regions, an off-centre peak and a central abundance drop. In both
the models and the data, the effects persist when two-temperature fits are
used (as they are in the inner regions when statistically required). It is
necessary to tie the abundances of the two components since there is not
enough information in the spectra to constrain them separately.

\subsection{Equivalent Width Effects}

The {\it ASCA} gradients were seen clearly in the equivalent width of the
iron K line. For the model spectra, however, excluding the iron L complex
by fitting only in the range 3-7\,keV produces different results. In this
case, whilst the abundance does rise initially in the central regions, the
large off-centre peak seen in Figure~1 never develops. Instead, the
abundance merely dies away in the inner regions at later times. Excluding
the iron K lines by fitting only to the spectral region 0.3--5\,keV
produces results very similar to those obtained using the full range.

\begin{figure}
  \plotfiddle{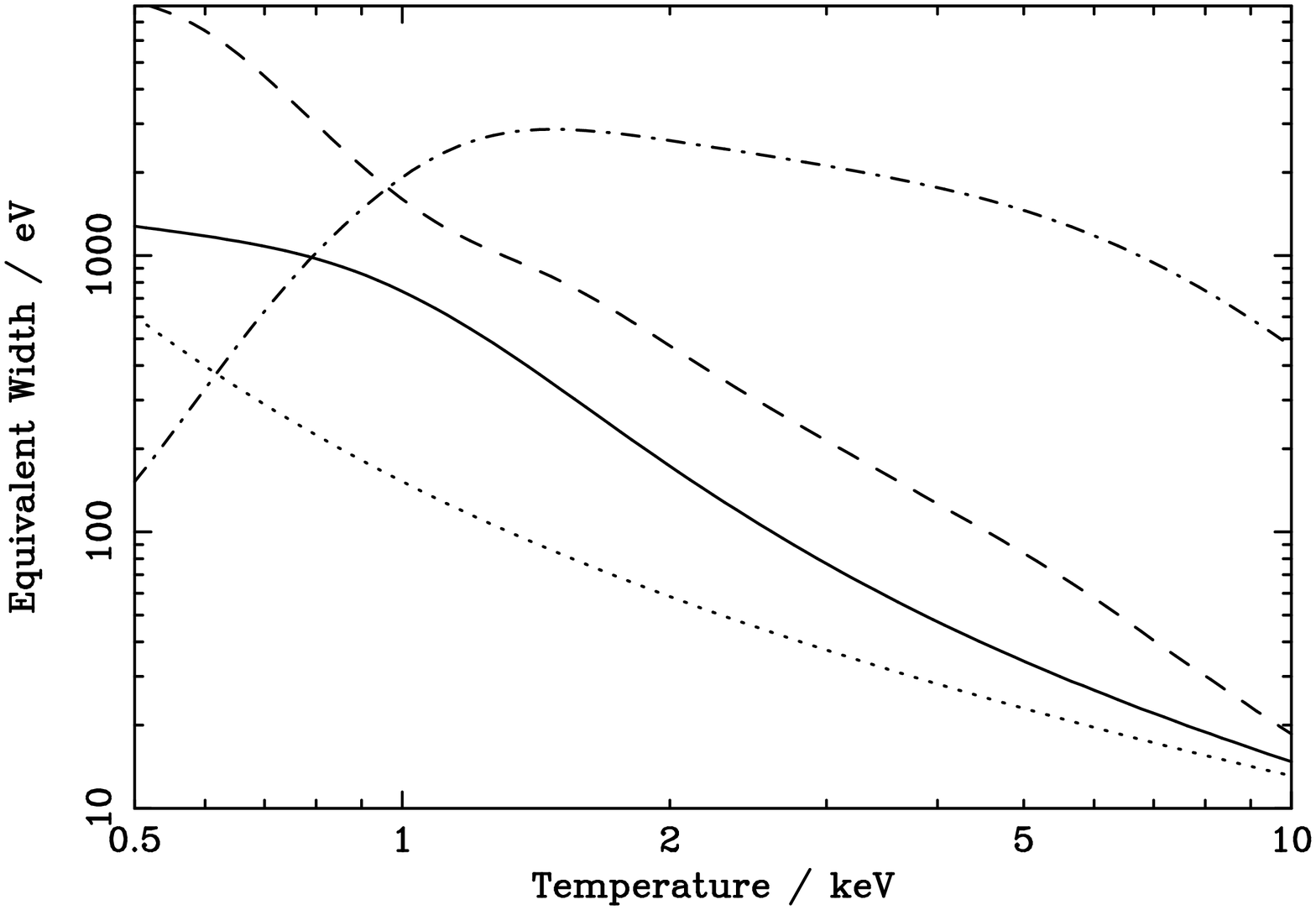}{235pt}{0}{50}{50}{-200}{280}
  \caption{Temperature dependence of some equivalent widths. Dot-dashed
    iron K; dashed iron L; dotted oxygen K; solid silicon K.}
\end{figure}

The discordancy between these two spectral features is caused by the
different temperature dependences of their equivalent widths, as
illustrated in Figure~2 (calculated from {\sc mekal} spectra). Starting
from high temperatures, the equivalent width of the iron L complex
increases strongly and monotonically with cooling. The K width on the other
hand increases only slightly, then falls off at lower temperatures. Thus
when the bimodal metallicity gas is allowed to cool, the strength of the L
complex relative to the continuum increases strongly, leading to pronounced
rises in the observed central abundance. The relative change in the iron K
lines is much less, so fitting exclusively to these spectral features
produces less of an effect. Thus small-scale variations cannot produce the
observed iron K gradients (but they are not inconsistent with the presence
of such gradients --- Morris \& Fabian, in preparation).

The equivalent widths of the silicon and oxygen K lines are also depicted
in Figure~2. Both show a monotonic increase on cooling, but the magnitude
of the effect is smaller for oxygen. Using the {\tt vmekal} model in {\it
XSPEC}, we may obtain separate abundance profiles for these elements.
Silicon shows a very similar profile to iron, but for oxygen whilst the
overall shape is the same, the peak abundance is significantly lower. Thus,
this is one way to produce an under-abundance of oxygen (but only in the
central regions of the ICM, and only with an accompanying gradient).

\section{Summary}

There is reason to think the metals in the ICM may not be smoothly
distributed. If this is the case, the cooling flow model predicts a reduced
flux of lines from low-temperature species, together with the appearance of
(artificial) abundance gradients in certain spectral lines, coupled with a
central drop in abundance. All these effects are seen in data (although of
course other explanations exist). The persistence of an abundance drop in
the centre of the ICM implies limits for the amount of convection and
mixing in these regions.

\end{document}